\documentclass[aps,prl,twocolumn,showpacs,10pt,floatfix,longbibliography]{revtex4-2}

\usepackage{graphicx}
\usepackage{subfigure}
\usepackage{amsmath}
\usepackage{amsfonts}

\usepackage{mathrsfs}
\usepackage{bbm}
\usepackage{subfigure}
\usepackage{xcolor}
\usepackage[colorlinks=true, urlcolor=blue, linkcolor=blue, citecolor=blue]{hyperref}

\usepackage[T1]{fontenc}

\usepackage[per-mode=symbol]{siunitx}
\usepackage{bm}
\usepackage[version=4]{mhchem} 

\usepackage{stackengine}[2013-09-11]

\usepackage{dsfont}

\AddToHook{cmd/appendix/before}{%
  \setcounter{equation}{0}%
  \setcounter{theorem}{0}%
}

\begin{document}

\title{Electric spin and valley Hall effects}

\author{W.~Zeng}
\email{zeng@ujs.edu.cn}
\affiliation{Department of Physics, Jiangsu University, Zhenjiang 212013, China}

\begin{abstract}

The electric Hall effect (EHE) is a newly identified Hall effect characterized by a perpendicular electric field inducing a transverse charge current in two-dimensional (2D) systems. Here, we propose a spin and valley version of EHE. We demonstrate that the transverse spin and valley currents can be generated in an all-in-one tunnel junction based on a buckled 2D hexagonal material in response to a perpendicular electric field, referred to as the electric spin Hall effect and electric valley Hall effect, respectively. These effects arise from the perpendicular-electric-field-induced backreflection phase of electrons in the junction spacer, independent of Berry curvature. The valley Hall conductance exhibits an odd response to the perpendicular electric field, whereas the spin Hall conductance shows an even one. The predicted effects can further enable the transverse separation of a pair of pure spin-valley-locked states with full spin-valley polarization while preserving time-reversal symmetry, as manifested by equal spin and valley Hall angles. Our findings present a new mechanism for realizing the spin and valley Hall effects and provide a novel route to the full electric-field manipulation of spin and valley degrees of freedom, with significant potential for future applications in spintronics and valleytronics.

\end{abstract}
\maketitle

\begin{figure}[tp]
\begin{center}
\includegraphics[clip = true, width =\columnwidth]{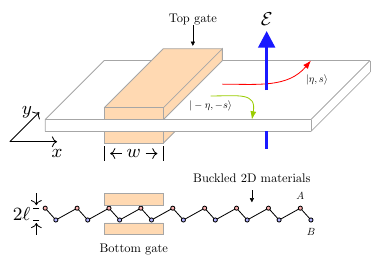}
\end{center}
\caption{ Schematic of the all-in-one tunnel junction based on a buckled 2D material (top panel) and of its side view (bottom panel), where the left and right electrode regions are separated by the central spacer region controlled by the top and bottom gates. A perpendicular electric field $\mathcal{E}$ (blue arrow) is applied on the right electrode region. The electrons with both opposite spin and valley are transversely separated, denoted by the red and green arrows. The red and blue circles in the bottom panel represent the $A$ and $B$ sublattices, respectively.  }
\label{fig:junction}
\end{figure}

\textit{Introduction.}---The ordinary Hall effect is a fundamental transport phenomenon in which a longitudinal current induces a transverse charge current under a perpendicular magnetic field \cite{d21672,hurd2012hall}. In ferromagnetic conductors, a transverse charge current can exist even without a magnetic field, known as the anomalous Hall effect \cite{RevModPhys.25.151,RevModPhys.82.1539,PhysRevLett.88.207208}. Beyond charge transport, a longitudinal current can also generate transverse spin and valley currents, giving rise to spin and valley Hall effects \cite{PhysRevLett.83.1834,valenzuela2006direct,PhysRevLett.108.196802,PhysRevLett.99.236809}. The extrinsic spin Hall effect originates from skew and side-jump scattering due to spin-orbit coupling \cite{PhysRevLett.112.066601,PhysRevLett.85.393}, whereas the intrinsic effect \cite{PhysRevLett.92.126603,PhysRevLett.97.126602,PhysRevLett.100.096401} arises entirely from spin-orbit coupling in the band structure, independent of scattering. Similarly, the intrinsic valley Hall effect arises from opposite Berry curvatures in different valleys \cite{PhysRevLett.99.236809,PhysRevB.103.195309,PhysRevB.102.035412}, whereas the extrinsic effect can be generated by tilted Dirac fermions \cite{PhysRevLett.131.246301,PhysRevLett.132.096302}.

A recent study \cite{5yxp-djy9} revealed that a perpendicular electric field can generate a transverse charge current in two-dimensional (2D) magnetic systems, known as the electric Hall effect. This naturally raises the question of whether a perpendicular electric field can also give rise to a transverse spin or valley current, thereby realizing a fully electric-field-controlled spin or valley Hall effect, which is of fundamental interest and holds significant potential for spintronic and valleytronic applications.

In 2D hexagonal materials such as silicene \cite{Chowdhury_2016,PhysRevLett.108.155501} and germanene \cite{PhysRevLett.116.256804,PhysRevB.84.195430}, the $A$ and $B$ sublattices are vertically displaced relative to the sheet plane; see Fig.\ \ref{fig:junction} (bottom). The buckled structure causes the sublattices to respond differently to an applied electric field. This unique property combined with controllable spin and valley degrees of freedom in such materials gives rise to a variety of intriguing spin-valley-dependent transport phenomena \cite{PhysRevB.92.195423,PhysRevB.104.205402,PhysRevB.91.125105,PhysRevB.88.085322,PhysRevB.87.241409}, such as spin-valley-polarized Andreev reflections \cite{PhysRevB.89.020504}, $0-\pi$ transitions \cite{PhysRevB.94.180505}, and the photoinduced topological phase transition \cite{PhysRevLett.110.026603}, suggesting that the buckled material serve as a promising platform for the design of electric-field-controlled spintronic and valleytronic devices.

In this Letter, we demonstrate that the transverse spin and valley currents can be generated in an all-in-one tunnel junction based on a buckled 2D hexagonal material in response to a perpendicular electric field $\mathcal{E}$. It is shown that the electrons in the junction spacer acquire an additional $\mathcal{E}$-induced backreflection phase, which depends on both the transverse momentum and spin-valley indices, leading to an asymmetric (skew) spin-valley-dependent transmission. Since $\mathcal{E}$ breaks the inversion symmetry ($\hat{\mathcal{P}}$) but preserves the time-reversal symmetry ($\hat{\mathcal{T}}$), the electrons with opposite spin and valley that form the Kramers pairs flow in opposite transverse directions, resulting in the Berry-curvature-free electric spin and valley Hall effects. We further demonstrate that the valley Hall conductance exhibits an odd response to $\mathcal{E}$, whereas the spin Hall conductance shows an even one. By tuning $\mathcal{E}$, a pair of pure spin-valley-locked states with full spin-valley polarization can be transversely separated without breaking $\hat{\mathcal{T}}$ symmetry, manifesting as equal-magnitude spin and valley Hall angles.

\begin{figure*}[tp]
\begin{center}
\includegraphics[clip = true, width =\linewidth]{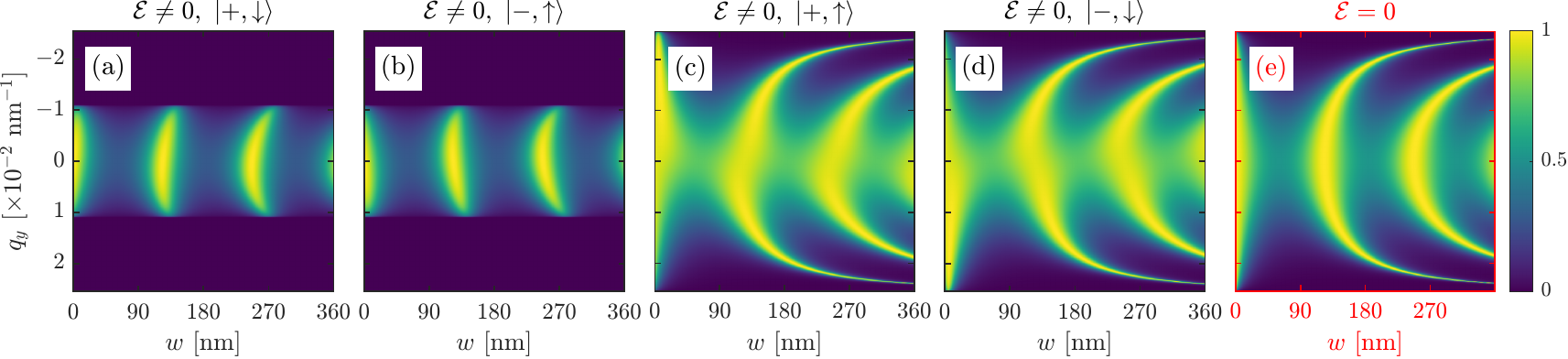}
\end{center}
\caption{Contour plot of the transmission probability $T_{\eta s}$ as a function of the junction length $w$ and the transverse wave number $q_y$ for $E=\SI{10}{\meV}$, $V=\SI{20}{\meV}$, and $\lambda_{\mathrm{SO}}=\SI{4}{\meV}$. (a)-(d) Spin- and valley-dependent transmission for $\mathcal{E}=\SI{22}{\milli\electronvolt\per\angstrom}$. The spin-valley channel labeled by $|\eta,s\rangle$ is indicated above the corresponding figure panel. (e) Transmission at $\mathcal{E}=0$.   }
\label{fig:tran}
\end{figure*}

\textit{Model and formalism.}---We consider an all-in-one tunnel junction based on a buckled 2D hexagonal material lying in the $x-y$ plane, as shown in Fig.\ \ref{fig:junction}. The longitudinal direction of the junction is along the $x$ axis. A $p$-doped spacer located at $0<x<w$, which is controlled by the voltage applied from the top and bottom gates, given by $\overline{V}=V\Theta(x)\Theta(w-x)$. A perpendicular electric field of magnitude $\mathcal{E}$ is applied in the right electrode region ($x>w$). The low-energy electron states can be described by the effective Hamiltonian \cite{PhysRevB.89.020504,PhysRevB.94.180505,PhysRevLett.110.026603}
\begin{align}
  \hat{\mathcal{H}}=-i\hbar v_F(\partial_x\hat{\tau}_x-\eta \partial_y\hat{\tau}_y)+\overline{\zeta}\hat{\tau}_z+\overline{V},
\end{align}
where $\hat{\tau}_i$ ($i=x,y,z$) is the Pauli matrix in the sublattice space, $v_F$ is the Fermi velocity, $\eta=+/-$ denotes the $K_+/K_-$ valleys of the electron. $\overline{\zeta}=\overline{\mathcal{E}}\ell-\eta s\lambda_{\mathrm{SO}}$ breaks $\hat{\mathcal{P}}$ symmetry and varies across different regions of the junction, where $\overline{\mathcal{E}}\ell$ is a staggered potential modulated by the perpendicular electric field in the buckled structure, with $2\ell$ denoting the distance between the two sublattice planes, and $\overline{\mathcal{E}}=\mathcal{E}\Theta(x-w)$ being the strength of the perpendicular electric field present only in the right electrode. $\lambda_{\mathrm{SO}}$ is the strength of the spin-orbit coupling and $s=\uparrow/\downarrow$ denotes the up/down spins of the electrons. 

In the presence of the perpendicular electric field, the four-fold degeneracy of the energy spectrum is lifted, resulting in 
\begin{align}
  \epsilon_\pm=\tau\sqrt{|\hbar v_F\vec{q}|^2+|\mathcal{E}\ell\pm\lambda_{\mathrm{SO}}|^2},
\end{align}
where $\vec{q}$ is the 2D momentum and $\tau=\pm1$ corresponds to the electrons in the conduction and valence bands, respectively. Here, $\epsilon_+$ is the band for electrons in states $|+,\downarrow\rangle$ and $|-,\uparrow\rangle$, while $\epsilon_-$ is the band for electrons in states $|+,\uparrow\rangle$ and $|-,\downarrow\rangle$. 

The scattering basis states are given by
\begin{align}
  \overline{\psi}_{\pm}(x,y)=\sqrt{|\overline{j}/j_1|}\begin{pmatrix}
    \pm\tau \overline{q}+i\eta q_y\\
    (E-\overline{Q})/\hbar v_F
  \end{pmatrix}e^{\pm i\tau \overline{q}x+iq_yy},\label{eq:basis}
\end{align}
where $\overline{j}=\langle\overline{\psi}_+|v_F\hat{\tau}_x|\overline{\psi}_+\rangle$ denotes the current flux in different regions of the junction, $j_1=\overline{j}|_{x<0}$, $\overline{Q}=\overline{V}+\overline{\zeta}$ is a coefficient determined by the junction parameters, the subscripts $+/-$ represent the scattering states propagating along the positive and negative directions of the $x$ axis, respectively. The longitudinal wave vector is given by $\overline{q}=(1/\hbar v_F)\sqrt{(E-\overline{V})^2-\overline{\zeta}^2-(\hbar v_Fq_y)^2}$, where $E$ is the incident energy and $q_y$ is the conserved transverse wave number. Note that since both $\overline{q}$ and $\overline{Q}$ take different values in different regions of the junction, the basis states in Eq.\ (\ref{eq:basis}) is a piecewise function. The scattering wave function is constructed from the basis states given in Eq.\ (\ref{eq:basis}), written as
\begin{align}
    \Psi(x,y)=\left\{
        \begin{array}{ll}
         \overline{\psi}_{+}(x,y)+r\overline{\psi}_{-}(x,y), & x<0, \\
          c_1\overline{\psi}_{+}(x,y)+c_2\overline{\psi}_{-}(x,y), & 0\leq x\leq w,\\
          t\overline{\psi}_{+}(x,y), & x>w,
        \end{array}
    \right.\label{eq:gf0}
\end{align}
where $c_{1}$ and $c_{2}$ are the linear combination coefficients of the scattering basis functions within the central spacer. $r$ and $t$ are the reflection and transmission amplitudes for the channel specified by the valley and spin indices $(\eta,s)$, respectively, which can be obtained by enforcing continuity of $\Psi(x,y)$ at $x=0$ and $x=w$. The net transmission probability for the $(\eta,s)$ channel can be obtained by $T_{\eta s}=|t|^2$.

\textit{Spin-valley-dependent asymmetric transmission.}---The numerical results are obtained using parameters relevant to silicene, where the intrinsic spin-orbit coupling is $\lambda_{\mathrm{SO}}=\SI{4}{meV}$, the distance between the two sublattice planes is $2\ell=\SI{0.46}{\angstrom}$, and the Fermi velocity is $v_F=\SI{5.5e5}{\meter\per\second}$.

Figures\ \ref{fig:tran}(a)-\ref{fig:tran}(d) show the transmission probability $T_{\eta s}$ as a function of both the transverse momentum $q_y$ and the spacer length $w$, at an incident energy of $E=\SI{10}{meV}$ and under a perpendicular electric field of $\mathcal{E}=\SI{22}{\milli\electronvolt\per\angstrom}$. Due to the spin-valley splitting induced by $\mathcal{E}$, electron tunneling depends on both spin and valley. The transmission of electrons in $\epsilon_+$ band (electrons in states $|+,\downarrow\rangle$ and $|-,\uparrow\rangle$) is restricted to a narrow range of the transverse momentum ($|q_y|\leq\sqrt{E^2-|\mathcal{E}\ell+\lambda_{\mathrm{SO}}|^2}/\hbar v_F\simeq\SI{1.08e-2}{\per\nano\meter}$), as shown in Figs.\ \ref{fig:tran}(a) and \ref{fig:tran}(b), which originates from the larger $\epsilon_+$-band gap compared to the gap associated with the incident states in the left electrode. In contrast, the transmission of the electrons in $\epsilon_-$ band (electrons in states $|+,\uparrow\rangle$ and $|-,\downarrow\rangle$) can transmit for all possible transverse wave number, as shown in Figs.\ \ref{fig:tran}(c) and \ref{fig:tran}(d). Remarkably, $T_{\eta s}$ is asymmetric with respect to $q_y$, \textit{i.e.}, $T_{\eta s}(q_y)\neq T_{\eta s}(-q_y)$. Nevertheless, the transmissions for the electrons related by $\hat{\mathcal{T}}$ symmetry are symmetric, satisfying $T_{\eta s}(q_y)=T_{-\eta -s}(-q_y)$. For $\mathcal{E}=0$, the spectrum is four-fold spin-valley degenerate, where the transmission cannot be resolved in terms of spin and valley degrees of freedom and exhibits a symmetric scattering pattern, as shown in Fig.\ \ref{fig:tran}(e).

The physical origin of this asymmetric transmission is attributed to the $\mathcal{E}$-induced additional backreflection phase acquired by the electrons within the central spacer. The net transmission amplitude across the spacer can be written in a general Fabry-P\'{e}rot form \cite{datta1997electronic,PhysRevLett.101.156804}
\begin{align}
  t=t_1t_2/(1-|r_1r_2|e^{i\varphi_{\mathrm{tot}}}),\label{eq:fb}
\end{align}
where $r_{1}$ ($t_1$) and $r_{2}$ ($t_2$) denote the reflection (transmission) amplitudes at the left and right interfaces of the spacer, respectively, satisfying $|t_{1,2}|^2+|r_{1,2}|^2=1$ due to the current conservation, and $\varphi_\mathrm{tot}$ is the total backreflection phase. One finds that
\begin{align}
  r_{l}=\varrho\times\frac{\rho_0-e^{(-1)^li\eta(\phi_0+\phi_l)}\rho_l}{e^{(-1)^li\eta\phi_0}\rho_0+e^{(-1)^li\eta\phi_l}\rho_l},
\end{align}
where $l=1,2$ denote the left and right interfaces of the spacer, respectively. $\varrho=1$ for $l=1$, and $\varrho=e^{2iq_0w}$ for $l=2$ with $q_0$ being the longitudinal wave vector in the spacer region. $\phi_0$, $\phi_1$, and $\phi_2$ are $q_y$-dependent transmission angles defined in the spacer, the left electrode, and the right electrode, respectively, while $\rho_{0}$, $\rho_{1}$, and $\rho_{2}$ are coefficients determined by the junction parameters in the corresponding regions. Both of them can be read off directly from the region-dependent parameters $\overline{\phi}=\arctan(q_y/\overline{q})$ and $\overline{\rho}=\sqrt{(E-\overline{V}+\overline{\zeta})/(E-\overline{Q})}$, respectively. The total backreflection phase is obtained by
\begin{align}
  \varphi_\mathrm{tot}=\arg(r_1)+\arg(r_2)=2q_0w+\varphi_G.
\end{align}
Here, $2q_0w$ represents a kinetic phase acquired by electrons moving through the spacer, while $\varphi_G$ is an additional phase acquired in the tunneling process caused by the perpendicular electric field, satisfying
\begin{align}
  \tan\varphi_G=\frac{2\kappa\rho_0[(\rho_0^2-1)E/\zeta_0+\rho_0^2+\rho_1\rho_2]}{(\rho_0^2-\rho_1^2)(\rho_0^2-\rho_2^2)\cos^2\phi_0+p_1p_2}\times \eta q_y\mathcal{E},\label{eq:pg}
\end{align}
where $\kappa=-\cos\phi_0/(E-\zeta_1)(E-\zeta_2)$, and $p_{l}=(\rho_0^2+\rho_l^2)\sin\phi_0+2\rho_0\rho_l\sin\phi_l$ with $l=1,2$. $\zeta_0$, $\zeta_1$, and $\zeta_2$ are the region-dependent parameters defined within the central spacer, the left electrode, and the right electrode, respectively, which can be directly read off from $\overline{\zeta}$. The presence of the additional phase $\varphi_G$ is associated with the finite solid angle on the Bloch sphere enclosed by the scattering states across the junction spacer, and is thus termed the geometric phase \cite{RevModPhys.64.51,RevModPhys.66.899,PhysRevB.89.155412}.

Most notably, $\varphi_G$ arises only when a nonzero perpendicular electric field is applied and flips its sign when $q_y$ is reversed \textit{i.e.}, $\varphi_G(-q_y)=-\varphi_G(q_y)$. The tunneling electrons with opposite $q_y$ acquire different backreflection phases, given by $2q_0w\pm\varphi_G$, respectively, leading to the transmission probability
\begin{align}
  |t(\pm q_y)|^2=[\beta_1+\beta_2\cos(2q_0w\pm\varphi_G)]^{-1},\label{eq:tqy}
\end{align}
where $\beta_1=(1+|r_1r_2|^2)/|t_1t_2|^2$ and $\beta_2=-2|r_1r_2|/|t_1t_2|^2$. Consequently, the nonzero $\varphi_G$ in Eq.\ (\ref{eq:tqy}), requiring a finite $\mathcal{E}$, underlies the asymmetric transmission. Note that the coefficients in Eq.\ (\ref{eq:pg}) are spin- and valley-dependent. Upon simultaneous reversal of the spin and valley indices, $\varphi_G$ reverses sign while preserving its magnitude, yielding symmetric scattering patterns between two electrons related by $\hat{\mathcal{T}}$ symmetry.

\begin{figure}[tp]
\begin{center}
\includegraphics[clip = true, width =\columnwidth]{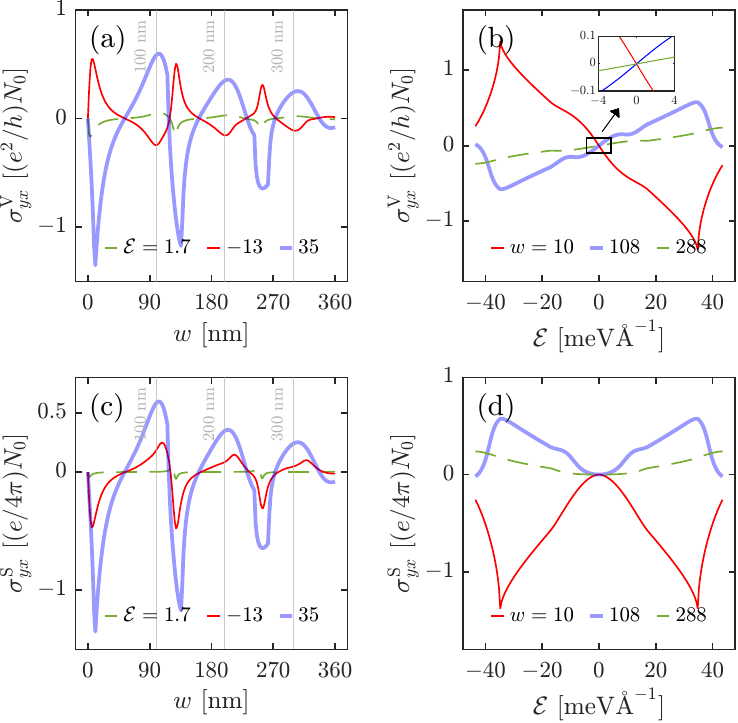}
\end{center}
\caption{[(a), (b)] Valley Hall conductance $\sigma^{\mathrm{V}}_{yx}$ versus $w$ and $\mathcal{E}$, respectively. [(c), (d)] Spin Hall conductance $\sigma^{\mathrm{S}}_{yx}$ versus $w$ and $\mathcal{E}$, respectively. $N_0=EW/\pi\hbar v_F$ is the number of the transverse modes with $W$ being the width of the junction. The incident energy is $E=\SI{8}{\meV}$. In the legend, the electric field is given in $\si{\milli\electronvolt\per\angstrom}$ and the length in $\si{nm}$. The remaining parameters are taken to be the same as Fig.\ \ref{fig:tran}. The inset in (b) shows a zoom-in of the boxed region.}
\label{fig:cond}
\end{figure}

\textit{Electric spin and valley Hall effects.}---This spin-valley-dependent asymmetric transmission induced by $\mathcal{E}$ may generate finite transverse spin and valley currents, quantified by the corresponding Hall conductance
\begin{align}
  \sigma_{yx}^{\mathrm{S}/\mathrm{V}}=\sum_{\eta s}\varsigma\times\sigma_{\eta s}^{\mathrm{T}},\label{eq:syx}
\end{align}
where $\varsigma=(\hbar/2e)s$ for $\sigma_{yx}^{\mathrm{S}}$ and $\varsigma=\eta$ for $\sigma_{yx}^{\mathrm{V}}$. $\sigma_{\eta s}^{\mathrm{T}}$ is the transverse conductance for $(\eta,s)$ channel, which is given by $\sigma_{\eta s}^{\mathrm{T}}=(e^2/h)\sum_{q_y}\xi(q_y)|t(q_y)|^2$ with $\xi(q_y)$ denoting the ratio of the transverse and longitudinal band velocities in the electrode.

The valley and spin Hall conductances ($\sigma_{yx}^{\mathrm{V}}$ and $\sigma_{yx}^{\mathrm{S}}$) as functions of the junction length $w$ for $E=\SI{8}{meV}$ and $V=\SI{20}{meV}$ are shown in Figs.\ \ref{fig:cond}(a) and \ref{fig:cond}(c), respectively. Both $\sigma_{yx}^{\mathrm{V}}$ and $\sigma_{yx}^{\mathrm{S}}$ show periodic oscillations with increasing $w$. These oscillations arise from the modulation of the kinetic phase $2q_0w$ encoded in the transmission probability, which is determined by the longitudinal wave vector $q_0$ in the junction spacer and independent of $\mathcal{E}$. The extrema appear at resonant lengths $w_n=n\pi/q_0\simeq n\times\SI{100}{nm}$ ($n\in\mathbb{Z}^+$), as indicated by the vertical lines in Figs.\ \ref{fig:cond}(a) and \ref{fig:cond}(c).

The dependence of $\sigma_{yx}^{\mathrm{V}}$ on $\mathcal{E}$ for different spacer lengths $w$ is presented in Fig.\ \ref{fig:cond}(b). A finite $\sigma_{yx}^{\mathrm{V}}$ arises at nonzero $\mathcal{E}$ and changes sign under $\mathcal{E}\rightarrow -\mathcal{E}$. This odd-$\mathcal{E}$ dependence can be understood from the geometric phase $\varphi_G$. A nonzero $\varphi_G$ induces transverse asymmetric transmission for electrons from different valleys. A positive $\sigma_{yx}^{\mathrm{V}}$ requires a finite valley current along the $+y$ axis, implying that the difference in transmission between $K_+$ and $K_-$ valleys is more pronounced in the $+q_y$ regime than in the $-q_y$ regime: 
\begin{align}
  \sum_s (T_{+s}-T_{-s})|_{q_y,\mathcal{E}}>\sum_s (T_{+s}-T_{-s})|_{-q_y,\mathcal{E}}.\label{eq:tes}
\end{align}
As shown in Eq.\ (\ref{eq:pg}), reversing both $q_y$ and $\mathcal{E}$ leaves $\varphi_G$ invariant, thus keeping $T_{\eta s}$ unchanged. Substituting $q_y\rightarrow -q_y$ and $\mathcal{E}\rightarrow-\mathcal{E}$ in Eq.\ (\ref{eq:tes}) yields $\sum_s (T_{+s}-T_{-s})|_{-q_y,-\mathcal{E}}>\sum_s (T_{+s}-T_{-s})|_{q_y,-\mathcal{E}}$, indicating that at $-\mathcal{E}$ the difference in transmission between $K_+$ and $K_-$ valleys is stronger in the $-q_y$ regime, which produces a valley current of equal magnitude along $-y$. Consequently, $\sigma_{yx}^{\mathrm{V}}$ changes sign upon reversal of $\mathcal{E}$, exhibiting an odd dependence $\sigma_{yx}^{\mathrm{V}}(-\mathcal{E})=-\sigma_{yx}^{\mathrm{V}}(\mathcal{E})$.

For weak perpendicular electric field ($|\mathcal{E}\ell|\ll\lambda_{\mathrm{SO}}$), the acquired geometric phase can be written as $\varphi_G\simeq \eta c_0\mathcal{E}$ to the first order of $\mathcal{E}$, where the coefficient $c_0$ is determined by Eq.\ (\ref{eq:pg}) and exhibits an odd dependence on $q_y$. Combining Eqs.\ (\ref{eq:tqy}) and (\ref{eq:syx}) shows that $\sigma_{yx}^{\mathrm{V}}(\mathcal{E})=\mu_0\cdot\mathcal{E}$, where $\mu_0=(e^2/h)\sum_{q_y\eta s}[c_0\xi(q_y)\beta_2\sin(2q_0w)]/[\beta_1+\beta_2\cos(2q_0w)]^2$ is a coefficient determined by the junction parameters, indicating that $\sigma_{yx}^{\mathrm{V}}$ exhibits a linear dependence on $\mathcal{E}$ in weak-$\mathcal{E}$ regime; as shown in the inset of Fig.\ \ref{fig:cond}(b), where $-\SI{4}{\milli\electronvolt\per\angstrom}<\mathcal{E}<\SI{4}{\milli\electronvolt\per\angstrom}$.

In contrast to $\sigma_{yx}^{\mathrm{V}}$, the response of $\sigma_{yx}^{\mathrm{S}}$ to $\mathcal{E}$ exhibits an even dependence, as shown in Fig.\ \ref{fig:cond}(d). For a transverse spin current along the $+y$ axis at a given $\mathcal{E}$, the transmission satisfies $\sum_\eta(T_{\eta\uparrow}-T_{\eta\downarrow})|_{q_y,\mathcal{E}}>\sum_\eta(T_{\eta\uparrow}-T_{\eta\downarrow})|_{-q_y,\mathcal{E}}$. From Eq.~(\ref{eq:pg}), it follows that for a given $q_y$, reversing both $\eta$ and $\mathcal{E}$ simultaneously does not change $\varphi_G$. As the spin Hall conductance sums over both valleys, swapping the valley indices $\eta$ leaves it unchanged, which gives rise to $\sum_\eta(T_{\eta\uparrow}-T_{\eta\downarrow})|_{q_y,-\mathcal{E}}>\sum_\eta(T_{\eta\uparrow}-T_{\eta\downarrow})|_{-q_y,-\mathcal{E}}$, implying that the direction of the spin current remains unchanged upon reversal of $\mathcal{E}$. Consequently, $\sigma_{yx}^{\mathrm{S}}$ shows an even dependence on $\mathcal{E}$, \textit{i.e.}, $\sigma_{yx}^{\mathrm{S}}(-\mathcal{E})=\sigma_{yx}^{\mathrm{S}}(\mathcal{E})$. $\sigma_{yx}^{\mathrm{S}}$ is absent at $\mathcal{E}=0$ and remains small in the weak-$\mathcal{E}$ regime, scaling as $\mathcal{O}(\mathcal{E}^2)$.

Electrons transversely separated onto the two edges of the junction are spin- and valley-polarized, with the valley-dependent spin polarization (or spin-valley polarization) defined as $P_{\pm}=(\sigma^{\mathrm{T}}_{\pm\uparrow}-\sigma^{\mathrm{T}}_{\pm\downarrow})/(\sigma^{\mathrm{T}}_{\pm\uparrow}+\sigma^{\mathrm{T}}_{\pm\downarrow})$ for $K_{\pm}$ valley. Electrons from opposite valleys always flow to separate edges, so $P_{\pm}$ also represent the spin-valley polarization of the states on the edges. By means of band engineering, two fully spin-valley-polarized states can be produced on each edge. For $\mathcal{E}>0$, the states $|+,\downarrow\rangle$ and $|-,\uparrow\rangle$ are fully blocked for incident energies within the band gap of $\epsilon_+$, \textit{i.e.}, $|\mathcal{E}\ell-\lambda_{\mathrm{SO}}|<E<|\mathcal{E}\ell+\lambda_{\mathrm{SO}}|$. Transmission occurs via the single Kramers pair $|+,\uparrow\rangle$ and $|-,\downarrow\rangle$, which are transversely separated, producing two pure spin-valley-locked states with full spin-valley polarization on each edge. Similarly, For $\mathcal{E}<0$, the fully spin-valley-polarized states $|-,\uparrow\rangle$ and $|+,\downarrow\rangle$ are transversely separated to the two edges for incident energies within the band gap of $\epsilon_-$, reversing the spin-valley polarization on each edge.

\begin{figure}[tp]
\begin{center}
\includegraphics[clip = true, width =\columnwidth]{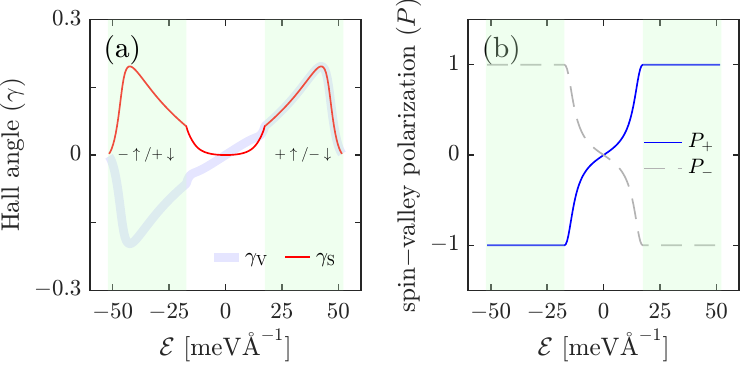}
\end{center}
\caption{Hall angle (a) and spin-valley polarization (b) versus $\mathcal{E}$ at $E=\SI{8}{meV}$, $\lambda_{\mathrm{SO}}=\SI{4}{\meV}$, and $w=\SI{198}{nm}$. }
\label{fig:ha}
\end{figure}

The generation of the fully spin-valley-polarized states is accompanied by an equal efficiency of the charge-to-spin and charge-to-valley conversions, which can be characterized by the spin and valley Hall angles, given by $\gamma_\mathrm{V}=\sigma^{\mathrm{V}}_{yx}/\sigma_{xx}$ and $\gamma_\mathrm{S}=(2e/\hbar)\sigma^{\mathrm{S}}_{yx}/\sigma_{xx}$, respectively. Here $\sigma_{xx}$ is the longitudinal conductance. The Hall angle and the spin-valley polarization as functions of $\mathcal{E}$ at $E=\SI{8}{meV}$ and $\lambda_{\mathrm{SO}}=\SI{4}{meV}$ are shown in Figs.\ \ref{fig:ha}(a) and \ref{fig:ha}(b), respectively. In the regime $\SI{4}{meV}<\mathcal{E}\ell<\SI{12}{meV}$ (corresponding to $\mathcal{E}\in[17.4,52]\ \si{\milli\electronvolt\per\angstrom}$), the fully spin-valley-polarized states $|+,\uparrow\rangle$ and $|-,\downarrow\rangle$ are generated and transversely separated to different edges. The valley Hall conductance is contributed by $\sigma_{yx}^{\mathrm{V}}=\sigma^\mathrm{T}_{+\uparrow}-\sigma^\mathrm{T}_{-\downarrow}$, while the spin Hall conductance is $\sigma_{yx}^{\mathrm{S}}=(\hbar/2e)(\sigma^\mathrm{T}_{+\uparrow}-\sigma^\mathrm{T}_{-\downarrow})=(\hbar/2e)\sigma_{yx}^{\mathrm{V}}$, leading to $\gamma_{\mathrm{V}}=\gamma_{\mathrm{S}}$ and $P_+=-P_-=1$; see the right green-shaded region in Figs.\ \ref{fig:ha}(a) and \ref{fig:ha}(b). In contrast, for $\mathcal{E}\in[-52,-17.4]\ \si{\milli\electronvolt\per\angstrom}$, the transmission is carried by the fully spin-valley-polarized electrons in the states $|-,\uparrow\rangle$ and $|+,\downarrow\rangle$, which are transversely separated, leading to $\sigma_{yx}^{\mathrm{S}}=-(\hbar/2e)\sigma_{yx}^{\mathrm{V}}=(\hbar/2e)(\sigma^{\mathrm{T}}_{-\uparrow}-\sigma^{\mathrm{T}}_{+\downarrow})$, and thus $\gamma_{\mathrm{V}}=-\gamma_{\mathrm{S}}$ and $P_-=-P_+=1$; see the left green-shaded region in Figs.\ \ref{fig:ha}(a) and \ref{fig:ha}(b). For $\mathcal{E}\in[-17.4,17.4]\ \si{\milli\electronvolt\per\angstrom}$, electrons from all spin-valley channels contribute to the transmission, which destroys the full spin-valley polarization on each edge, thereby resulting in $\gamma_\mathrm{V}\neq\gamma_\mathrm{S}$ and $|P_{\pm}|<1$. These pure spin-valley-locked states, separated onto opposite edges of the junction and exhibiting full spin-valley polarization, can be readily manipulated by an external electric field without breaking $\hat{\mathcal{T}}$ symmetry, and may be of potential benefit to fields such as spin-valley-based quantum computing~\cite{Rohling_2012} and the preparation of spin-valley entangled states \cite{PhysRevB.108.235137,duprez2024spin}.

Distinct from the recently reported electric (charge) Hall effect \cite{5yxp-djy9}, which is determined intrinsically by the Berry curvature polarization and Berry curvature polarizability, the predicted electric spin and valley Hall effects originate from an extrinsic mechanism associated with the phase-coherent tunneling. Indeed, this proposed extrinsic mechanism in our model can also generate a transverse charge current [\textit{i.e.}, the electric (charge) Hall effect] by breaking the $\hat{\mathcal{T}}$ symmetry, which can be realized either through proximity-induced ferromagnetism or antiferromagnetism \cite{PhysRevB.110.035426,PhysRevB.94.180505}, or by applying off-resonant circularly polarized light in the spacer region \cite{PhysRevLett.110.026603,PhysRevB.98.125424}. The simple geometry of the junction allows for easy experimental implementation. The predicted effect is realized in a tunnel junction and is thus related to the previously reported tunneling Hall effect \cite{PhysRevLett.115.056602,PhysRevLett.131.246301,PhysRevLett.117.166806,PhysRevLett.110.247204}. However, the underlying physics of our model differs from that of previous studies. The tunneling anomalous and spin Hall effects \cite{PhysRevLett.115.056602} originate from the $\hat{\mathcal{T}}$ symmetry mismatch between a magnetic electrode and a nonmagnetic barrier, while the tunneling valley Hall effect is attributed to the mismatch of the tilt-induced anisotropic Fermi surfaces \cite{PhysRevLett.131.246301}. On the contrary, the predicted electric spin and valley Hall effects are realized in a tunnel junction with preserved $\hat{\mathcal{T}}$ symmetry and isotropic Fermi surfaces and most importantly are induced by an externally applied perpendicular electric field, distinguishing it from earlier studies.

\textit{Conclusions.}---In summary, we have predicted the electric spin and valley Hall effects in an all-in-one tunnel junction based on a buckled 2D hexagonal material, where both the transverse spin and valley currents can be produced by applying a perpendicular electric field $\mathcal{E}$. These effects are independent of Berry curvature and are attributed to the $\mathcal{E}$-induced backreflection phase acquired by the tunneling electrons in the spacer. Using the parameters of an existing buckled material, we demonstrate that the transverse valley Hall conductance exhibits an odd response to $\mathcal{E}$, whereas the spin Hall conductance shows an even one. By tuning the perpendicular electric field, a pair of pure spin-valley-locked states with full spin-valley polarization can be transversely separated while preserving $\hat{\mathcal{T}}$ symmetry, manifesting as equal-magnitude spin and valley Hall angles. Our work offers a new mechanism for realizing the spin and valley Hall effects and a novel approach to full electric-field manipulation of the spin and valley degrees of freedom, which holds significant promise for future applications in spintronics and valleytronics.

\textit{Acknowledgements.}---This work was supported by the National Natural Science Foundation of China (Grant No.\ 12504052), the Natural Science Foundation of Jiangsu Province (Grant No.\ BK20250838), and the Natural Science Foundation of the Jiangsu Higher Education Institutions of China (Grant No.\ 25KJB140004).

\end{document}